\documentclass[twocolumn,showpacs,aps,prl,superscriptaddress]{revtex4}

\usepackage{graphicx}
\usepackage{dcolumn}
\usepackage{epsfig}
\usepackage{amsmath}

\newcommand{\BaBarYear}    {03}
\newcommand{\BaBarNumber}  {041}

\newcommand{\SLACPubNumber} {10240}

\input pubboard/babarsym
\RequirePackage{xspace}

\newcommand{\pvec}{{\bf p}}

\newcommand{\acp}{\ensuremath{\calA_{ch}}}
\newcommand{\calB}{\ensuremath{{\cal B}}}

\newcommand{\bfemsix}{${\cal B}(10^{-6}$)}

\newcommand{\DE}{\ensuremath{\Delta E}}

\newcommand{\mres}{\ensuremath{m_{\rm res}}}
\newcommand{\xf}{\ensuremath{{\cal F}}}
\newcommand{\hel}{\ensuremath{{\cal H}}}
\newcommand{\thetaT}{\ensuremath{\theta_{\rm T}}}
\newcommand{\costhr}{\ensuremath{\cos\thetaT}}

\newcommand\etal{{\it et al.}}
\newcommand{\half}{\ensuremath{{1\over2}}}

\newcommand{\bma}[1]{\boldmath{$#1$}}
\newcommand{\msp}{\phantom{-}}

\newcommand{\bfig}{\begin{figure}[htbpc!]}
\newcommand{\efig}{\end{figure}}
\newcommand\bef{\begin{figure}}
\newcommand\edf{\end{figure}}
\newcommand\dbline{\noalign{\vskip 0.10truecm\hrule}\noalign{\vskip 2pt}\noalign{\hrule\vskip 0.10truecm}}
\providecommand{\tbline}{\noalign{\vskip 0.05truecm\hrule\vskip0.05truecm}}

\newcommand\beq{\begin{equation}}
\newcommand\eeq{\end{equation}}
\newcommand\bear{\begin{array}}
\newcommand\enar{\end{array}}
\newcommand\beqa{\begin{eqnarray}}
\newcommand\eeqa{\end{eqnarray}}
\newcommand\ben{\begin{enumerate}}
\newcommand\een{\end{enumerate}}

\newcommand{\UfourS}{\ensuremath{\Upsilon(4S)}}

\newcommand{\etagg}{\ensuremath{\eta_{\gaga}}}
\newcommand{\etappp}{\ensuremath{\eta_{3\pi}}}
\newcommand{\etatogg}{\ensuremath{\eta\ra\gaga}}
\newcommand{\etatoppp}{\ensuremath{\eta\ra\pi^+\pi^-\pi^0}}

\newcommand{\etapepp}{\ensuremath{\etapr_{\eta\pi\pi}}}
\newcommand{\etaptoepp}{\ensuremath{\etapr\ra\eta\pip\pim}}
\newcommand{\etaprg}{\ensuremath{\etapr_{\rho\gamma}}}
\newcommand{\etaptorg}{\ensuremath{\etapr\ra\rho^0\gamma}}

\newcommand{\omtoppp}{\ensuremath{{\omega\ra\pip\pim\piz}}}

\newcommand{\kzs}{\ensuremath{\KS}}

\newcommand{\fetapip}{\ensuremath{\eta\pi^+}}
\newcommand{\etapip}{\ensuremath{\Bp\ra\fetapip}}
\newcommand{\Betapip}{\ensuremath{\calB(\etapip)}}
\newcommand{\retapip}{\ensuremath{xx^{+xx}_{-xx}\pm xx}}

\newcommand{\Aetapip}{\ensuremath{xx^{+xx}_{-xx}\pm xx}}
\newcommand{\AetapipBG}{\ensuremath{xx^{+xx}_{-xx}\pm xx}}
\newcommand{\setapip}{\ensuremath{xx}}
   \newcommand{\fetaggpip}{\ensuremath{\eta_{\gaga} \pip}}
   
   \newcommand{\fetappppip}{\ensuremath{\eta_{3\pi} \pip}}

\newcommand{\fetaKp}{\ensuremath{\eta K^+}}
\newcommand{\etaKp}{\ensuremath{\Bp\ra\fetaKp}}
\newcommand{\BetaKp}{\ensuremath{\calB(\etaKp)}}
\newcommand{\retaKp}{\ensuremath{xx^{+xx}_{-xx}\pm xx}}

\newcommand{\AetaKp}{\ensuremath{xx^{+xx}_{-xx}\pm xx}}
\newcommand{\AetaKpBG}{\ensuremath{xx^{+xx}_{-xx}\pm xx}}
\newcommand{\setaKp}{\ensuremath{xx}}
   \newcommand{\fetaggKp}{\ensuremath{\eta_{\gaga} \Kp}}
   
   \newcommand{\fetapppKp}{\ensuremath{\eta_{3\pi} \Kp}}

\newcommand{\fetaKz}{\ensuremath{\eta\Kz}}
\newcommand{\etaKz}{\ensuremath{\Bz\ra\fetaKz}}
\newcommand{\BetaKz}{\ensuremath{\calB(\etaKz)}}
\newcommand{\retaKz}{\ensuremath{xx^{+xx}_{-xx}\pm xx}}

\newcommand{\uletaKz}{\ensuremath{xx}}

\newcommand{\setaKz}{\ensuremath{xx}}

   \newcommand{\fetaggKz}{\ensuremath{\eta_{\gaga}\Kz}}
   
   \newcommand{\fetapppKz}{\ensuremath{\eta_{3\pi}\Kz}}

\newcommand{\fetappip}{\ensuremath{\etapr\pip}}
\newcommand{\etappip}{\ensuremath{\Bp\ra\fetappip}}
\newcommand{\Betappip}{\ensuremath{\calB(\Bp\ra\etapr \pip)}}
\newcommand{\retappip}{\ensuremath{xx^{+xx}_{-xx} \pm xx}}

\newcommand{\uletappip}{\ensuremath{xx}}

\newcommand{\setappip}{\ensuremath{xx}}
   \newcommand{\fetapepppip}{\ensuremath{\etapr_{\eta\pi\pi} \pi^+}}
   
   \newcommand{\fetaprgpip}{\ensuremath{\etapr_{\rho\gamma} \pi^+}}
   \newcommand{\etaprgpip}{\ensuremath{\Bp\ra\fetaprgpip}}

\newcommand{\fomegapip}{\ensuremath{\omega\pi^+}}
\newcommand{\omegapip}{\ensuremath{\Bp\ra\fomegapip}}
\newcommand{\Bomegapip}{\ensuremath{\calB(\omegapip)}}
\newcommand{\romegapip}{\ensuremath{xx \pm xx\pm xx}}

\newcommand{\Aomegapip}{\ensuremath{xx\pm xx \pm xx}}

\newcommand{\AomegapipBG}{\ensuremath{xx\pm xx \pm xx}}
\newcommand{\somegapip}{\ensuremath{xx}}

\newcommand{\fomegaKp}{\ensuremath{\omega K^+}}
\newcommand{\omegaKp}{\ensuremath{\Bp\ra\fomegaKp}}
\newcommand{\BomegaKp}{\ensuremath{\calB(\omegaKp)}}
\newcommand{\romegaKp}{\ensuremath{xx\pm xx\pm xx}}

\newcommand{\AomegaKp}{\ensuremath{xx\pm xx \pm xx}}
\newcommand{\AomegaKpBG}{\ensuremath{xx\pm xx \pm xx}}

\newcommand{\somegaKp}{\ensuremath{xx}}

\newcommand{\fomegaKz}{\ensuremath{\omega K^0}}
\newcommand{\omegaKz}{\ensuremath{\Bz\ra\fomegaKz}}
\newcommand{\BomegaKz}{\ensuremath{\calB(\omegaKz)}}
\newcommand{\romegaKz}{\ensuremath{xx^{+xx}_{-xx}\pm xx}}

\newcommand{\somegaKz}{\ensuremath{xx}}

\providecommand{\bfemsix}{${\cal B} (10^{-6})$}
\providecommand{\msp}{\phantom{$-$}}
\renewcommand{\retapip}{\ensuremath{5.3\pm1.0\pm0.3}}
\renewcommand{\Aetapip}{\ensuremath{-0.44\pm0.18\pm0.01}}
\renewcommand{\AetapipBG}{\ensuremath{-0.003\pm0.008}}
\renewcommand{\setapip}{\ensuremath{7.9}}
\renewcommand{\retaKp}{\ensuremath{3.4\pm0.8\pm0.2}}
\renewcommand{\AetaKp}{\ensuremath{-0.52\pm0.24\pm0.01}}
\renewcommand{\AetaKpBG}{\ensuremath{-0.010\pm0.011}}
\renewcommand{\setaKp}{\ensuremath{6.1}}
\renewcommand{\retaKz}{\ensuremath{2.9\pm1.0\pm 0.2}}
\renewcommand{\uletaKz}{\ensuremath{5.2}}
\renewcommand{\setaKz}{\ensuremath{3.3}}

\renewcommand{\retappip}{\ensuremath{2.7\pm1.2\pm 0.3}}
\renewcommand{\uletappip}{\ensuremath{4.5}}

\renewcommand{\setappip}{\ensuremath{3.4}}

\renewcommand{\romegapip}{\ensuremath{5.5\pm0.9\pm0.5}}
\renewcommand{\Aomegapip}{\ensuremath{0.03\pm0.16\pm0.01}}
\renewcommand{\AomegapipBG}{\ensuremath{0.012\pm0.006}}
\renewcommand{\somegapip}{\ensuremath{9.1}}
\renewcommand{\romegaKp}{\ensuremath{4.8\pm0.8\pm0.4}}
\renewcommand{\AomegaKp}{\ensuremath{-0.09\pm0.17\pm0.01}}
\renewcommand{\AomegaKpBG}{\ensuremath{-0.003\pm0.009}}
\renewcommand{\somegaKp}{\ensuremath{10.0}}
\renewcommand{\romegaKz}{\ensuremath{5.9^{+1.6}_{-1.3}\pm0.5}}
\renewcommand{\somegaKz}{\ensuremath{7.5}}

\begin{document}

\preprint{\babar-PUB-\BaBarYear/\BaBarNumber} 
\preprint{SLAC-PUB-\SLACPubNumber} 

\begin{flushleft}
\babar-PUB-\BaBarYear/\BaBarNumber\\
SLAC-PUB-\SLACPubNumber\\
\end{flushleft}




\title{
 \large \bf\boldmath Observation of \omegaKz, \etapip, and \etaKp\ and Study of Related Decays 
}

%
\author{B.~Aubert}
\author{R.~Barate}
\author{D.~Boutigny}
\author{F.~Couderc}
\author{J.-M.~Gaillard}
\author{A.~Hicheur}
\author{Y.~Karyotakis}
\author{J.~P.~Lees}
\author{V.~Tisserand}
\author{A.~Zghiche}
\affiliation{Laboratoire de Physique des Particules, F-74941 Annecy-le-Vieux, France }
% ...
\collaboration{The \babar\ Collaboration}
\noaffiliation

\date{\today}

\begin{abstract}
We present measurements of branching fractions and charge
asymmetries for seven $B$-meson decays with an $\eta$, \etapr\ or $\omega$
meson in the final state.
The data sample corresponds to 89 million \BB\ pairs produced from \epem\ 
annihilation at the \UfourS\ resonance.  We measure the following branching 
fractions in units of $10^{-6}$:
$\Betapip=\retapip$, $\BetaKp=\retaKp$, $\BetaKz=\retaKz$ 
($<\uletaKz$, 90\% C.L.), $\Betappip=\retappip$ ($<\uletappip$, 90\% C.L.),
$\Bomegapip=\romegapip$, $\BomegaKp=\romegaKp$, and $\BomegaKz=\romegaKz$.  
The charge asymmetries are $\acp(\etapip)=\Aetapip$, 
$\acp(\etaKp)=\AetaKp$, $\acp(\omegapip)=\Aomegapip$ and $\acp(\omegaKp)=\AomegaKp$.
\end{abstract}


\maketitle

We report results of measurements of $B$ decays to the charmless
final states \fetaKz, \fetapip, \fetaKp, \fetappip, \fomegaKz,
\fomegapip, and \fomegaKp\ \cite{CC}.  Only the last two of these
decays have been observed previously \cite{CLEOomega, BABARomega,  BELLEomega}.
Measurements of the related $\B\ra\etapr K$ decays were published
recently \cite{etaprPRL}.
Charmless decays with kaons are usually expected to be dominated by $b\ra s$ loop
(``penguin") amplitudes, while $b\ra u$ tree transitions are typically larger for
the decays with pions.  However the $\B\ra\eta K$ decays are especially 
interesting since they are suppressed relative to the abundant $\B\ra\etapr K$ 
decays due to destructive interference between two penguin amplitudes
\cite{Lipkin}.  Thus the CKM-suppressed 
$b\ra u$ amplitudes may interfere significantly with the suppressed penguin 
amplitudes.  This tree-penguin interference may lead to large direct \CP\ 
violation in the \fetaKp\ decay as well as \fetapip, and \fetappip\ 
\cite{directCP}; numerical estimates have been provided in a few
cases \cite{acpgrabbag}.  We search for 
such direct \CP\ violation by measuring the charge asymmetry 
$\acp \equiv (\Gamma^--\Gamma^+)/(\Gamma^-+\Gamma^+)$ in the rates
$\Gamma^\pm=\Gamma(B^\pm\ra f^\pm)$, for each observed charged final
state $f^\pm$.

Charmless $B$ decays are becoming useful to test the
accuracy of theoretical predictions such as QCD factorization \cite{BN}.
Phenomenological fits to the branching fractions and charge asymmetries
can be used to understand the importance of tree and penguin contributions 
and may even provide sensitivity to the CKM angle $\gamma$ \cite{chiang}.

The results presented here are based on data collected
with the \babar\ detector~\cite{BABARNIM}
at the PEP-II asymmetric $e^+e^-$ collider~\cite{pep}
located at the Stanford Linear Accelerator Center.  An integrated
luminosity of 81.9~fb$^{-1}$, corresponding to 
$88.9\pm1.0$ million \BB\ pairs, was recorded at the $\Upsilon (4S)$
resonance (center-of-mass energy $\sqrt{s}=10.58\ \gev$).

Charged particles from the \epem\ interactions are detected, and their
momenta measured, by a combination of a vertex tracker (SVT) consisting
of five layers of double-sided silicon microstrip detectors, and a
40-layer central drift chamber, both operating in the 1.5-T magnetic
field of a superconducting solenoid. We identify photons and electrons 
using a CsI(Tl) electromagnetic calorimeter (EMC).
Further charged particle identification (PID) is provided by the average energy
loss ($dE/dx$) in the tracking devices and by an internally reflecting
ring imaging Cherenkov detector (DIRC) covering the central region.

We select $\eta$, $\etapr$, $\omega$, \KS, and $\piz$ candidates through the 
decays \etatogg\ (\etagg), \etatoppp\ (\etappp), 
\etaptoepp\ (\etapepp), \etaptorg\ (\etaprg),
\omtoppp, $\rho^0\ra\pip\pim$, $\kzs\ra\pip\pim$, and $\piz\ra\gaga$.  We
make the following requirements on the invariant mass (in \mev) of their final
states: $490< m_{\gaga}<600$ for \etagg, $520<m_{\pi\pi\pi}< 570$ for
\etappp, $910<(m_{\eta\pi\pi},m_{\rho\gamma})<1000$ for \etapr,
$735<m_{\pi\pi\pi}<825$ for $\omega$, 
$510 < m_{\pi\pi} <1070$ for $\rho^0$,
and $120 < m_{\gaga} < 150$ for \piz.  For
\kzs\ candidates we require $488 < m_{\pi\pi} < 508$,
the three-dimensional flight distance from the event primary vertex to be 
greater than 2 mm, and the angle between flight and momentum vectors, in the 
plane perpendicular to the beam direction, to be less than 40 mrad.

We make several PID requirements to ensure the identity of the pions and kaons.
Secondary tracks in \etappp, \etapr, and $\omega$ candidates must have DIRC,
$dE/dx$, and EMC outputs consistent with pions.  For the \Bp\ decays to an 
$\eta$ or $\omega$ meson and a charged pion or kaon, the latter (primary) track 
must have an associated DIRC signal with a Cherenkov angle within $3.5$ standard 
deviations ($\sigma$) of the expected value for either a $\pi$ or $K$ hypothesis.

A $B$-meson candidate is characterized kinematically by the energy-substituted 
mass $\mes=\lbrack{(\half s+\pvec_0\cdot\pvec_B)^2/E_0^2-\pvec_B^2}\rbrack^\half$
and energy difference $\DE = E_B^*-\half\sqrt{s}$, where the subscripts $0$ and
$B$ refer to the initial \UfourS\ and to the $B$ candidate, respectively,
and the asterisk denotes the \UfourS\ frame. 
The resolution on \DE\ (\mes) is about 30 MeV ($3.0\ \mev$). 
We require $|\DE|\le0.2$ GeV and $5.2\le\mes\le5.29\ \gev$.

Backgrounds arise primarily from random combinations in $\epem\ra\qqbar$
events. We reject these by using the angle
\thetaT\ between the thrust axis of the $B$ candidate in the \UfourS\
frame and that of the 
rest of the charged tracks and neutral clusters in the event.
The distribution of $|\costhr|$ is
sharply peaked near $1.0$ for combinations drawn from jet-like \qqbar\
pairs, and nearly uniform for $B$-meson decays.  We require $|\costhr|<0.9$, 
for all modes except the high-background \etaprgpip\ decay, where we
determine that the sensitivity is maximal for a 0.65 requirement.
We also use, in the fit 
described below, a Fisher discriminant \xf\ that combines four variables: the
angles with respect to the beam axis of the $B$ momentum and $B$ thrust axis 
(in the \UfourS\ frame), and the zeroth and second angular moments $L_{0,2}$ 
of the energy flow about the $B$ thrust axis.  The moments are defined by
$ L_j = \sum_i p_i\times\left|\cos\theta_i\right|^j,$
where $\theta_i$ is the angle with respect to the $B$ thrust axis of
track or neutral cluster $i$, $p_i$ is its momentum, and the sum
excludes the $B$ candidate.

For the \etatogg\ modes we use additional event-selection criteria to 
reduce \BB\ backgrounds from several charmless final states. 
We reduce background from $B\ra\pip\piz$, $\Kp\piz$, and $\Kz\piz$ by 
rejecting \etagg\ candidates that share a photon with any \piz\ candidate having
momentum between 1.9 and 3.1 GeV/c in the \UfourS\ frame.  Additionally, we
require $E_{\gamma} < 2.4$ GeV to suppress background from $B\ra\Kstar\gamma$
and related radiative-penguin decays.  From Monte Carlo (MC) simulation 
\cite{geant} we estimate that the residual charmless \BB\ background is 
negligible for all decays except those with \etatogg\ and \etaptorg, where we 
include in the fit described below a \BB\ component (which is less than 0.5\% of
the total sample in all cases).

We obtain yields and \acp\ from extended unbinned 
maximum-likelihood fits, with input observables \DE, \mes, \xf, $\mres$ (the 
mass of the $\eta$, \etapr, or $\omega$ candidate), 
for the $\omega$ decays, $\hel\equiv |\cos{\theta_H}|$, and for charged modes 
the PID variable $S_{\pi,K}$.  The helicity angle $\theta_H$ is 
defined as the angle, measured in the $\omega$ rest frame, between the normal 
to the $\omega$ decay plane and the flight direction of the $\omega$.  
We incorporate PID information by using $S_\pi$ ($S_K$), 
the number of standard deviations between the measured Cherenkov angle
and the expectation for pions (kaons).

For each event $i$, hypothesis $j$ (signal, continuum background, 
\BB\ background), and flavor (primary \pip\ or \Kp)
$k$, we define the  probability density function (PDF)
\begin{eqnarray}
{\cal P}^i_{jk} &=&  {\cal P}_j (\mes^i) {\cal  P}_j (\DE^i_k) 
 { \cal P}_j(\xf^i) {\cal P}_j (\mres^i) \nonumber \\
 && \times\left[{\cal P}_j
(S^i_k)\right]\left[{\cal P}_j (\hel^i)\right].
\end{eqnarray}
The terms in brackets for $S$ and \hel\ pertain to modes with charged
track or $\omega$ daughters, respectively.  The absence of correlations
among observables in the background ${\cal P}^i_{jk}$ is confirmed in the
(background-dominated) data samples entering the fit.  For the signal
component, we correct for the effect of the neglect of small correlations 
(see below).  The likelihood function is
\begin{equation}
{\cal L} = \exp{(-\sum_{j,k} Y_{jk})}
\prod_i^{N}\left[\sum_{j,k} Y_{jk} {\cal P}^i_{jk}\right]\,,
\end{equation}
where $Y_{jk}$ is the yield of events of hypothesis $j$ and flavor $k$
found by maximizing \calL, and $N$ is the number of events in the sample.  
  
We determine the PDF parameters from simulation for the
signal and \BB\ background components, and from (\mes,\,\DE) sideband
data for continuum 
background.  We parameterize each of the functions ${\cal P}_{\rm sig}(\mes),\ 
{\cal  P}_{\rm sig}(\DE_k),\ { \cal P}_j(\xf),\ { \cal P}_j(S_k)$ and the 
peaking components of ${\cal P}_j(\mres)$ with either a Gaussian, the sum of
two Gaussians or an asymmetric Gaussian function as required to describe the 
distribution.  Slowly varying distributions (mass, energy or helicity-angle 
for combinatoric background) are represented by linear or quadratic dependencies.
The peaking and combinatoric components of the $\omega$ mass spectrum
each have their own $\hel$ shapes.  The combinatoric background in \mes\
is described by the function $x\sqrt{1-x^2}\exp{\left[-\xi(1-x^2)\right]}$,
with $x\equiv2\mes/\sqrt{s}$ and parameter $\xi$.
Large control samples of $B$ decays to charmed final states of similar 
topology are used to verify the simulated resolutions in \DE\ and \mes.  
Where the control data samples reveal differences from MC in mass or energy
offset or resolution, we shift or scale the resolution used in the likelihood fits.

\begin{figure}[!tb]
\vspace{0.5cm}
 \includegraphics[angle=0,width=\linewidth]{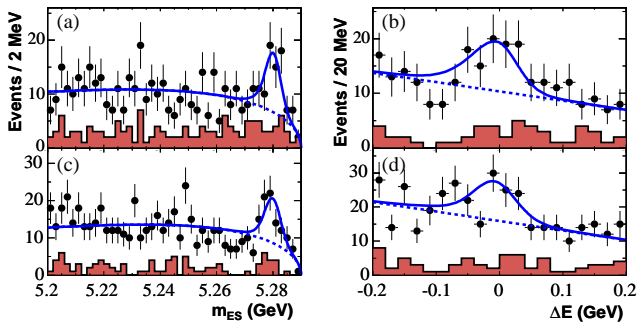}
 \caption{\label{fig:projMbDE_eta}
Projections of the $B$ candidate \mes\ and \DE\ for (a, b) \etapip, and
(c, d) \etaKp.  Points with errors represent data, 
shaded histograms the \etatoppp\ subset, solid curves the full fit functions,
 and dashed curves the background functions (the peaking \BB\ background
component is negligible).
These plots are made with a requirement on the likelihood and thus do not 
show all events in the data samples.
  }
\end{figure}

\begin{figure}[!htb]
\vspace{0.5cm}
 \includegraphics[angle=0,width=\linewidth]{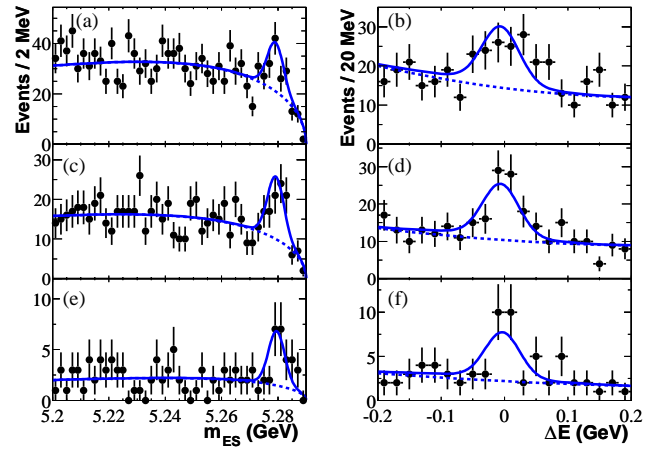}
 \caption{\label{fig:projMbDE_omega}
Projections of the $B$ candidate \mes\ and \DE\ for (a, b) \omegapip; 
(c, d) \omegaKp; and (e, f) \omegaKz. Points with errors represent data, 
solid curves the full fit functions, and dashed curves the background functions.
These plots are made with a requirement on the likelihood and thus do not 
show all events in the data samples.
  }
\end{figure}

In Table \ref{tab:results} we show for each decay mode the measured branching 
fraction, together with the quantities entering into its computation.
Typically seven parameters of the background PDF are free in the fit, 
along with signal 
and background yields, and for charged modes the signal and background \acp.
For calculation of branching fractions, we assume that the decay rates 
of the \UfourS\ to \BpBm\ and \BzBzb\ are equal.
For the $\eta$ and \etapr\ decays, we combine results from the two decay
channels by adding the values of $-2\ln{\cal L}$, taking proper account
of the correlated and uncorrelated systematic errors.
The estimated purity is the ratio of the signal yield to the effective
background plus signal; we estimate the effective background by 
taking the square of the uncertainty of the signal yield as the sum of 
effective background plus signal.
In Figs.\ \ref{fig:projMbDE_eta} and \ref{fig:projMbDE_omega} we show 
projections onto \mes\ and \DE\ after requiring $S_{\pi,K}\lsim 2$ 
[for (a)--(d)] and requiring that the signal likelihood (computed ignoring 
the PDF associated with the variable plotted) exceeds a mode-dependent
threshold.

\begin{table*}[btp]
\caption{
Signal yield, estimated purity $P$, detection
efficiency $\epsilon$, daughter branching fraction product,
significance (including systematic uncertainties), measured branching fraction,
background ($\acp^{qq}$) and signal (\acp) charge asymmetries for each mode.
For \etaKz\ and \etappip, the 90\% C.L. upper limit is also given.}
\label{tab:results}
\begin{tabular}{lcccccccc}
\dbline
Mode	      & Yield	   &$P$ (\%)&$\epsilon$ (\%)&$\prod\calB_i$ (\%)& Signif. &  \bfemsix 	& $\acp^{qq}$ & \acp  \\
\tbline
~~\fetappppip &   $28^{+10}_{-9}$ &30&23&23&4.4&$5.6^{+2.1}_{-1.8}$&$-0.004\pm0.010$&$-0.52\pm0.31$ \\
~~\fetaggpip  &      $59\pm14$    &31&31&39&6.6&    $5.2\pm1.3$    &$-0.001\pm0.011$&$-0.41\pm0.22$\\
\bma{\fetapip}&                   &  &  &  &\bma{\setapip}&\bma{\retapip}& \AetapipBG &\bma{\Aetapip} \\
~~\fetapppKp  &   $15^{+8}_{-7}$  &24&23&23&2.6&$3.1^{+1.7}_{-1.5}$&$-0.008\pm0.016$&$-0.43\pm0.51$ \\
~~\fetaggKp   &     $38\pm11$     &33&23&39&5.3&    $3.5\pm1.1$    &$-0.011\pm0.016$&$-0.55\pm0.26$\\
\bma{\fetaKp} &                   &  &  &  &\bma{\setaKp} &\bma{\retaKp} & \AetaKpBG &\bma{\AetaKp} \\
~~\fetapppKz  &$2.6^{+4.1}_{-3.1}$&20&23& 8&0.8&$1.8^{+2.9}_{-2.2}$&   &    \\
~~\fetaggKz   &$8.6^{+4.8}_{-3.8}$&47&24&14&3.2&$3.2^{+1.8}_{-1.4}$&   &    \\
\bma{\fetaKz} &                   &  &  &  &\bma{\setaKz} &\bma{\retaKz}{$~(<\uletaKz)$}&   &    \\
~~\fetapepppip &  $17^{+7}_{-6}$  &38&28&17&3.9&$3.8^{+1.7}_{-1.4}$&   &    \\
~~\fetaprgpip &  $-4^{+11}_{-9}$  &   &17&30&   &$-0.8^{+2.4}_{-2.0}$&   &    \\
\bma{\fetappip} &                 &  &  &  &\bma{\setappip}&\bma{\retappip}{$~(<\uletappip)$}&   &    \\
\bma{\fomegapip}&   $101\pm17$    &37&23&89&\bma{\somegapip}&\bma{\romegapip}&{\msp\AomegapipBG}&\bma{\msp\Aomegapip}\\
\bma{\fomegaKp}&     $83\pm14$    &39&22&89&\bma{\somegaKp}&\bma{\romegaKp}&\AomegaKpBG&\bma{\AomegaKp} \\
\bma{\fomegaKz}& $33^{+9}_{-8}$   &51&20&31&\bma{\somegaKz}&\bma{\romegaKz}&   &    \\
\dbline
\end{tabular}
\vspace{-5mm}
\end{table*}

The statistical error on the signal yield and \acp\ is taken as the change in 
the central value when the quantity $-2\ln{\cal L}$ increases by one 
unit from its minimum value. The significance is taken as the square root 
of the difference between the value of $-2\ln{\cal L}$ (with systematic 
uncertainties included) for zero signal and the value at its minimum.  
For \fetaKz\ and \fetappip\ we quote a 90\% confidence level (C.L.) upper 
limit, taken to be the branching 
fraction below which lies 90\% of the total of the likelihood integral 
in the positive branching fraction region.
For the charged modes we also give the charge asymmetry \acp .

Most of the yield uncertainties arising from lack of knowledge of the PDFs have 
been included in the statistical error since most background 
parameters are free in the fit.  Varying the signal PDF parameters within 
their estimated uncertainties, we estimate the uncertainties in the signal 
PDFs to be 1--3 events.  We verify the validity of the fit procedure and
PDF shapes by demonstrating that the 
likelihood of each fit is consistent with the distribution found in simulation. 

Uncertainties in our knowledge of the efficiency, found from auxiliary 
studies, include 0.8$N_t$\%, 2.5$N_\gamma$\%, and 3\%\ for a
\KS\ decay, where $N_t$ and $N_\gamma$ are the number of signal tracks
and photons, respectively.
Our estimate of the $B$ production systematic error is 1.1\%.  
The neglect of correlations among observables in the fit can cause a 
systematic bias; the correction for this bias ($<10$\% in all cases) and
assignment of systematic uncertainty (1--5\%), is determined from 
simulated samples with varying background populations.  Published data
\cite{PDG2002}\ provide the uncertainties in the $B$-daughter product branching 
fractions (1\%).
Selection efficiency uncertainties are 1\% (3\%\ in \etaprgpip) for
\costhr\ and $\sim$1\% for PID. 
Using several large inclusive kaon and $B$-decay samples, we find a
systematic uncertainty for \acp\ of 1.1\%\ due mainly to the dependence of
reconstruction efficiency on the charge of the high momentum charged track.
The values of $\acp^{qq}$ (see Table \ref{tab:results}) provide
confirmation of this estimate.

In conclusion, we find significant signals for five $B$-meson
decays.  The measured branching fractions, and for the \Bpm\ modes
the charge asymmetries, are given in Table
\ref{tab:results}.  These are the first charge asymmetry measurements for 
the decays \etapip\ and \etaKp, since these modes along with \omegaKz\
have not been 
observed previously.  We quote 90\% C.L. upper limits for the \etaKz\ and 
\etappip\ branching fractions,
where the significances are only $\setaKz\sigma$ and $\setappip\sigma$, 
respectively.  All branching fraction and charge asymmetry measurements
are consistent with, but more precise than, previous
measurements \cite{CLEOomega, BABARomega,  BELLEomega, CLEOeta}.
Though uncertainties are large, the values of \acp\ for the two decays with 
$\omega$ mesons are small as expected theoretically; the consistencies with zero
asymmetry for \etapip\ (\etaKp) are 2.4$\sigma$ (2.1$\sigma$).
These are channels in which large asymmetries may be anticipated \cite{directCP}.

We are grateful for the excellent luminosity and machine conditions
provided by our \pep2\ colleagues, 
and for the substantial dedicated effort from
the computing organizations that support \babar.
The collaborating institutions wish to thank 
SLAC for its support and kind hospitality. 
This work is supported by
DOE
and NSF (USA),
NSERC (Canada),
IHEP (China),
CEA and
CNRS-IN2P3
(France),
BMBF and DFG
(Germany),
INFN (Italy),
FOM (The Netherlands),
NFR (Norway),
MIST (Russia), and
PPARC (United Kingdom). 
Individuals have received support from the 
A.~P.~Sloan Foundation, 
Research Corporation,
and Alexander von Humboldt Foundation.

\end{document}